\documentstyle[preprint,aps,amssymb,epsf]{revtex}
\tightenlines
\newcommand{\beq}{\begin{equation}}
\newcommand{\eeq}{\end{equation}}
\newcommand{\bea}{\begin{eqnarray}}
\newcommand{\eea}{\end{eqnarray}}
\begin{document}

\title{Quasi-one and two-dimensional transitions of gases adsorbed on nanotube
bundles}

\author{S. M. Gatica$^1$, M. J. Bojan$^2$, G. Stan$^3$, and M. W. Cole$^4$}

\address{$^1$ Departamento de F\'{\i}sica, Facultad de Ciencias Exactas y Naturales, Universidad de Buenos Aires, 1428, Buenos Aires,Argentina. \\ 
$^2$ Departments of Chemistry Pennsylvania State University University Park, PA 16802, USA\\ 
$^3$ Institute for Physical Science and Technology and\\ 
Department of Chemical Engineering, University of Maryland, College Park, MD 20742, USA.\\
$^4$ Departments of Physics, Pennsylvania State University University Park, PA 16802, USA }

\date{\today}
\maketitle

\begin{abstract}

Grand canonical Monte Carlo simulations have been performed to
determine the adsorption behavior of Ar and Kr atoms on the exterior
surface of a rope (bundle) consisting of many carbon nanotubes. The
computed adsorption isotherms reveal phase transitions associated with
the successive creation of quasi-one dimensional lines of atoms near
and parallel to the intersection of two adjacent nanotubes.
\end{abstract}

\section{Introduction}

Much attention has focused recently on the problem of gas adsorption
within bundles (``ropes") of carbon nanotubes
\cite{Simonyan,Inoue,Gao,Teizer,Weber,Zambano,Yin,Kuznetsova,Williams,Dresselhaus,Stan1,Cole1,Stan2}.
For small atoms and molecules, both the cylindrical spaces within
individual tubes and the interstitial channels between tubes are
regions of very attractive potential energy for adsorption. Larger
molecules experience a similar attraction within the tubes (but not
within the interstitial region, unless swelling occurs). In either
case, there occurs significant uptake, within these respective
regions, at pressures P below saturated vapor pressure P$_0$ (svp). It
was noted recently by Williams and Eklund \cite{Williams} that finite
bundles of carbon nanotubes ought to exhibit additional, significant
adsorption on the external surface of the bundles.  In the
thermodynamic limit of an infinite array, such external adsorption
represents a negligible fraction of the total. Typical finite bundles,
however, have radius R$\sim 100$ {\AA} and are expected to manifest
significant external adsorption. Indeed, experiments of Talapatra {\it
et al.} \cite{Talapatra} have been interpreted as providing evidence
of such external adsorption in the case of CH$_4$ molecules, Ne and Xe
atoms.

In this paper we formulate a simple model of this system with which we
explore the phenomenon of external adsorption. We suppose that the
nanotube bundle of Fig. \ref{fig:bundle} is our subject of
investigation.  This assumption provides us with a normalization
factor (essentially a surface/volume ratio) required for a specific
prediction of the adsorbate mass per mass of substrate (and nothing
more). Any alternative geometry requires a ``renormalization" factor,
discussed below. Our actual calculations of external adsorption are
performed with a simplified abstraction of figure
\ref{fig:bundle}. Specifically, the model employs a periodic, planar
array of parallel cylinders representing the nanotubes. This planar
model should be an adequate approximation to the real situation (tubes
at the perimeter of a rope) in an experimental situation for which R
greatly exceeds the radius ($\sim$7 {\AA}) of an individual tube. Our
method of evaluating the adsorption is computer simulation, using the
grand canonical Monte Carlo (GCMC) method. With this well developed
technique we are able to demonstrate the existence of several phase
transitions in the film. These include both two-dimensional (2D)
transitions (analogous to layering transitions on flat surfaces) and
quasi-1D transitions (analogous to ones on similar quasi-1D
geometries, such as a surface facet cut close to that of a low index
surface).  \cite{weiss}

The outline of this paper is the following. The following section
describes our model potential and computational method. Section
\ref{sec:results} describes our results.  In section
\ref{sec:discussion}, we describe implications for experiments
studying such systems.

\section{ Method}
\label{sec:method}

As in most adsorption studies on graphite or carbon nanotubes, the
adsorption potential assumed here is a pairwise sum of two-body interactions
$U({\bf x})$ between a molecule and the nanotube's carbon atoms
\cite{many-body,kostov}:
\begin{equation}
V({\bf r}) = \sum_i U ({\bf r} - {\bf R_i})
\end{equation}
where {\bf r} is the position of the molecule and {\bf R$_i$} is the
position of a C atom.  The pair potential is assumed to be isotropic
and of Lennard-Jones form: $U (r) = 4 \epsilon
[(\sigma/r)^{12}-(\sigma/r)^6]$.  Although we neglect here the effect
of anisotropy for adsorption on graphite \cite{anisotropy}, the
potential considered is suitable for our qualitative study. The LJ
parameters for the Ar-C and Kr-C interaction are obtained with
semiempirical combining rules from the corresponding LJ parameters :
\cite{steele1,steele2,scoles,ihm}:
\begin{eqnarray}
\sigma_{g C} &=& \frac{\sigma_{gg} + \sigma_{C C}}{2}\\
\epsilon_{g C} &=& \sqrt{\epsilon_{gg}\ \epsilon_{C C}} \nonumber
\end{eqnarray}
The parameter values used in this study are listed in the Table \ref{table:LJ}.

Another simplifying assumption employed here is the replacement of
discrete carbon atoms by a continuous cylindrical sheet of matter;
this should not drastically affect the adsorbate's behavior except
when the adjacent tubes are in perfect registry, an unlikely
situation.  The resulting potential at distance $r$ from the nanotube
axis, outside of the tube, is \cite{stan}:
\begin{equation}
V(r;R)=3\ \pi\ \theta\ \epsilon\ \sigma {R}  \biggl[ \frac{21}{32}
\biggl(\frac{\sigma}{r}\biggr)^{11}   M_{11}(x)
- \biggl( \frac{\sigma}{r}\biggr)^5 M_5(x) \biggr]  
\end{equation}
where $\theta = 0.38$ \AA$^2$ is the surface density of C atoms and
$R$ is the radius of the nanotube. We use the integrals
\begin{equation}                                                            
M_{n}(x) = \int_0^{\pi} d\varphi \frac{1}{(1 + x^2 - 2 x
 cos\varphi)^{n/2}}.  
\end{equation}                                                              
Finally, the adsorption potential on the external surface of the
nanotube bundle is obtained by summing up the interactions of the
molecule with all the nanotubes in the bundle.

We perform GCMC simulations of Ar and Kr interacting with the outer
wall of a bundle of (10,10) tubes lined up to create a ``flat" surface
with grooves.  In the GCMC simulations, the chemical potential,
temperature and volume are held constant while the number of particles
varies. This technique is standard and has been described in textbooks
\cite{frenkel,allen}, so only the details will be given here.  Three
kinds of moves are performed: displacement of a molecule, creation of
a particle and deletion of a particle.  In the original method of
Norman and Filinov,\cite{norman} these moves were done in equal
proportion (33\% each). In our simulations however, the percentage of
attempted creation and deletions was set at 40\% each while 20\% of
the attempted moves were displacements. This was done because the
acceptance rate for displacement moves was typically much higher than
the acceptance rates for the creation/deletion moves. By increasing
the percentage of attempts, the total number of creations and
deletions that are accepted is increased, thus reducing the
computation time somewhat. It is still necessary to perform an
increased number of moves in the transition regions and in the high
density regions to insure good thermodynamic averages.  For a single
isotherm point typically, $3\times 10^6$ moves were performed to
equilibrate the system and 10$^6$ moves were used for data gathering.

The Ar-Ar and Kr-Kr pair interactions were taken as a 12-6
Lennard-Jones potential.  In representing the interaction of the gas
with the bundle of nanotubes, only one tube was modeled, with the
groove in the center of the simulation cell and the apex of the tube
at the periodically replicated boundary. Since the distance between
the axis of two neighboring nanotubes is 17 {\AA}, one of the cell
boundaries (the x direction) was fixed at this length. The tube length
(y-dimension) was set at 10$\sigma_{gg}$ and periodic boundary
conditions were used in these two dimensions. The height of the
simulation box (z-direction) was set at 40 {\AA} where a reflecting
surface was used.  Figure \ref{fig:potential} depicts the adsorption
potential for the case of an Ar atom in the unit cell.

In Table \ref{table:energies} we compare values of the well depth
found here with those found for Ne, Ar and Kr atoms in related
geometries: the interstitial channel, inside the nanotubes and on the
surface of graphite. It is seen that Ne experiences its most
attractive potential in the interstitial channels, while the other
(larger) atoms find the external groove to be most attractive. In the
latter cases, the well is nearly twice as attractive as the well on
the graphite surface. This can be understood from the fact that the
most favorable position in the groove case corresponds to the optimal
distance from each of the contributing tubes and each tube contributes
a well depth equal to 85\% of that on planar graphite, so that the
groove is 70\% more attractive than graphite.  This situation is
similar to recent measurements by Talapatra {\it et al.}
\cite{Talapatra}.  Their reported binding energies for Ne, Xe and
CH$_4$ adsorbed in closed-ended nanotubes are about 75\% larger than
on planar graphite.

Simulations of Ar interacting with the bundle surface at temperatures
ranging from 30K to 68K were performed.  The effect of molecular size
was tested by varying the size ($\sigma$) of Argon (ranging from 3.6
to 4.0 {\AA}) without varying the interaction strength. These results
will be referred to as ``artificial Argon''. Also, simulations of the
adsorption of Kr on this surface for temperatures ranging from T = 57K
to 97K were performed to look at the effect of increased interaction
strength.

\section{Results}
\label{sec:results}

Fig. \ref{fig:isoar} presents results for the adsorption of Ar at low
temperature (less than or of order one-half of the $\epsilon_{ArAr}$
parameter). None of these results exhibits any evidence of
hysteresis. The coverage in this and subsequent figures is expressed
in two alternative forms on the ordinate scales. On the left appears a
number equal to the number of atoms per unit length of the simulation
cell, which is 10$\sigma_{ArAr}$=34 {\AA}. For a single line of atoms
at close packing, this corresponds to about nine atoms per 1d
chain. Our checks show that the simulation results do not differ
significantly if we increase this periodicity length. The right scale
expresses the coverage in adsorbate moles per gram of carbon. That
calculation is carried out by assuming the geometry to be the
hexagonal structure of Fig. \ref{fig:bundle} (which contains N$_t$=37
nanotubes and N$_g$ =18 grooves on its perimeter). If the bundle size
or shape is different, one merely corrects the axis label by
multiplying the right scale's values by a factor (37 N$_g$)/(18
N$_t$).

One observes in Fig. \ref{fig:isoar} that the isotherm at T=30 K
consists of a series of plateaus, separated by steps. Each plateau
represents a region of stability (over a range of P) of a particular
structure, which one observes in the density plots shown in
Fig. \ref{fig:dens}. This behavior is analogous to the stepped
isotherms seen in adsorption on flat surfaces, a manifestation of
layering transitions on microscopically flat surfaces.

At very low P, significant adsorption occurs only in a ``groove"
formed along the contact line of two nanotubes. Below the first
plateau, this isotherm coincides with that $\mu_{1d}$(P,T) predicted
for a purely one-dimensional system, with an additive constant V$_1$
due to the potential provided by neighboring nanotubes and a small,
slowly varying term $\delta\!\mu_T$ arising from adatom motion
transverse to the groove:
\beq
\mu= V_1 + \mu_{1d}(P,T) +\delta\!\mu_T
\eeq

\beq 
\beta\,\delta\!\mu_T= ln [ \lambda^2/ ( \pi\langle r^2\rangle )
].  
\eeq 
Here 1/$\beta$ = k$_B$T and $\langle r^2\rangle$ = 2/($\beta$
k) is the mean square displacement perpendicular to the channel,
expressed in terms of the transverse force constant k, and
$\lambda=\sqrt{(2\pi\beta\hbar^2)/m}$ is the de Broglie thermal
wavelength. The values of V$_1$ and other potential energies at
locations of specific 1d lines of atoms are presented in Table
\ref{table:v}.

Above the plateau associated with saturation (complete filling) of the
groove, there occurs a coverage jump near P = 10$^{-14}$ atm. (at
T=30K).  This corresponds to the appearance of a well-defined
``three-channel" regime, the density of which appears in Fig. 4a. Upon
further increasing the pressure by a factor of $\sim$ 10, the coverage
jumps by a factor of two, to a regime in which the surface is covered
by a striped monolayer film, seen in Fig. 4b. A further increase in P
by a factor of 3,000 results in a transition to a bilayer film,
depicted in Fig. 4c. This film grows continuously thereafter, since
this is a strongly wetting situation.

The pressures at which these jumps occur can be predicted rather
accurately with simple model estimates. For example, the threshold at
T=30 K for forming the three-channel state is estimated to be:
\beq 
\mu_3 = V_3 - 4 \epsilon_{Ar}+ \delta\!\mu_T = -1352 K 
\eeq 
The factor of four is based on an assumed 4-fold coordination in the
``new'' pair of channels (beyond the original one), plus an external
field V$_3$ contributed by atoms in the original channel of
adatoms. The preceding estimate is close to the value found in the
simulation ($\mu_{3}$= - 1270 K) for the onset of the three-channel
state. Encouraged by this degree of consistency, we estimate the
chemical potential threshold for the monolayer, for which there are
six channels present:
\beq
\mu_{6}=  - 3 \epsilon_{Ar} + V_6 = -1155 K
\eeq
Here the factor of three is derived from one-half of the 6-fold
coordination.  This numerical result is close to that found in the
simulation (near P = 10$^{-13}$ atm, $\mu_{6}$ = - 1040 K).

In Fig. \ref{fig:isokr} we present the adsorption isotherms for Kr at
temperatures ranging from 43K to 97K. In the isotherm at T=43K we find
the same transitions than for Ar at 30K. (Notice that the reduced
temperatures are the same, i.e. 43K/$\epsilon_{KrKr}$=30K/$\epsilon_{ArAr}=0.25$)

One of the interesting features of these isotherms is the role played
by size commensuration in determining the monolayer transition on this
surface comprised of nanotubes. This effect is analogous to the role
played by the molecular diameter in determining the presence, or
absence, of epitaxial phases in monolayer phases on planar
surfaces.\cite{sander,bruch} Figure \ref{fig:arti} presents isotherms
obtained for a set of ``artificial Argon'' systems; these gases differ
from one another in the values of the $\sigma$ parameter, as
indicated. As a result of this variation, the monolayer transition is
seen to change in a discontinuous way. For small values of $\sigma$,
the monolayer consists of 6 lines of atoms.  Above a threshold value,
$\sigma$=3.8{\AA}, the monolayer consists of only 5 lines of
atoms. This behavior is an expected steric effect. It is interesting
to see that there exists a transitional value, $\sigma\sim$3.6{\AA} ,
for which ambiguity is present in the isotherm at this monolayer
transition (near P=10$^{-9}$ atm).

The isotherms for Ar and Kr presented earlier both correspond to the
small atom case and therefore we find that the monolayer film consists
of six parallel lines of atoms. On the basis of the artificial systems
reported in Fig. \ref{fig:arti}, we expect the monolayer of Xe (not
studied here) to correspond to five parallel lines.

\section{Discussion}
\label{sec:discussion}

These calculations predict the existence of transitions between
unusual phases of matter. The first transition (as a function of P) is
that between a one-dimensional fluid within a groove and a set of
three parallel lines of fluid. Subsequent transitions occur to a
monolayer and bilayer phase. Such transitions are possible due to the
interaction between one group (e.g., three lines) of particles and
other groups which are present in the system, which is taken to be
periodic. The observation that a doubling of the transverse cell
dimension does not affect the results significantly suggests that
these results are present in the infinite system (as is the case in
our recent study of the interstitial phase's condensation transition
\cite{Cole1}). Of course, one recognizes that the actual system under
investigation is not infinite; it is a quasi-cylindrical bundle of
(approximately parallel) nanotubes which is quite finite in transverse
dimension.  Hence one expects a rounding of the vertical jumps shown
here.  The degree of rounding depends on the deviation from our
assumptions. Once that is known, more realistic simulations can be
carried out.

This research has been supported by the National Science Foundation's
Focussed Research Group program. We are grateful to Moses Chan, Peter
Eklund, John Lewis, Paul Sokol, Bill Steele and Keith Williams for
stimulating discussion.

\newpage

\begin{table}[h]
\caption{Parameters of the LJ interaction for Ar and Kr.\protect\cite{steele2,watts}}
\vspace*{0.2 in}
\begin{tabular} {ccccc} 
Gas&$\epsilon_{gg}$(K)&$\sigma_{gg}$(\AA)& $\epsilon_{gC}$(K)&$\sigma_{gC}$(\AA)
\\ \hline
Ar&       120&  3.4 & 57.9&3.4\\
Kr&   171&  3.6&  69.2&3.5
\end{tabular}
\label{table:LJ}
\end{table}

\newpage

\begin{table}[h]
\caption{Characteristic well depths of Ar Kr and Ne in the external
surface (ext), the interstitial channel (IC), inside the nanotubes
(NT) and on flat graphite (gr). \protect\cite{Stan2}}
\vspace*{0.2 in}
\begin{tabular} {ccccc} 
Gas&$V_{min}^{ext}$(K)&$V_{min}^{IC}$(K)&$V_{min}^{NT}$(K)&$V_{min}^{gr}$(K)
\\ \hline
Ar&  -1607&6&-1426&-965 \\
Kr&  -2025&2048&-1836&-1220\\
Ne& -725&-1018&-600&-431\\
\end{tabular}
\label{table:energies}
\end{table}

\newpage

\begin{table}[h]
\caption{Potential energies of Ar atoms located in the central
channel (1) and the following channels numbered from the center to the
right (see fig. \ref{fig:dens}b).}
\vspace*{0.2 in}
\begin{tabular}{ccccc} 
channel&1&2&3&4
\\ \hline
V(K)&-1607&-872&-799&-795
\end{tabular}
\label{table:v}
\end{table}

\newpage

\begin{figure}[ht]
\caption{ A model nanotube bundle possessing 37 tubes and 18 external grooves. }
\label{fig:bundle}
\end{figure}

\begin{figure}[ht]
\caption{ Depiction of one unit cell of the one-dimensionally periodic
line of nanotubes assumed in the simulations. The contours correspond
to constant potential energy values V/$\epsilon_{ArC}$=
-25,-20,-15,-10,-5,-1 from darker to lighter. The dashed lines
correspond to the cylindrical nanotube surface.}
\label{fig:potential}
\end{figure}

\begin{figure}[ht]
\caption{ Adsorption isotherms for Ar at various temperatures,
indicated by the symbols in the figure. The left ordinate label
$\langle N\rangle$ is defined in text. The right scale, as described
in text, assumes a bundle having the structure depicted in figure
\ref{fig:bundle}. }
\label{fig:isoar}
\end{figure}

\begin{figure}[ht]
\caption{Density of Ar atoms as a function of the coordinates
perpendicular to the axis of the bundle. The temperature is 30K and
the pressure is a) P=0.36 10$^{-13}$atm ($\langle N\rangle$=27.3),
b)P=0.4 10$^{-12}$atm ($\langle N\rangle$=54.0), c)P=0.59 10$^{-9}$atm
( $\langle N\rangle$=99.5). The contours show constant density values
of 2{\AA}$^{-2}$ (thick) and 0.2{\AA}$^{-2}$ (thin).  }
\label{fig:dens}
\end{figure}

\begin{figure}[ht]
\caption{Adsorption isotherms for Kr at several temperatures indicated
in the legend.}
\label{fig:isokr}
\end{figure}

\begin{figure}[ht]
\caption{Adsorption isotherms for four gases having $\epsilon_{gg}=
\epsilon_{ArAr}$ and $\sigma_{gg}$ indicated in the inbox. The
temperature is 40K.}
\label{fig:arti}
\end{figure}

\newpage
\begin{figure}[ht]
\epsfysize=5.in \epsfbox{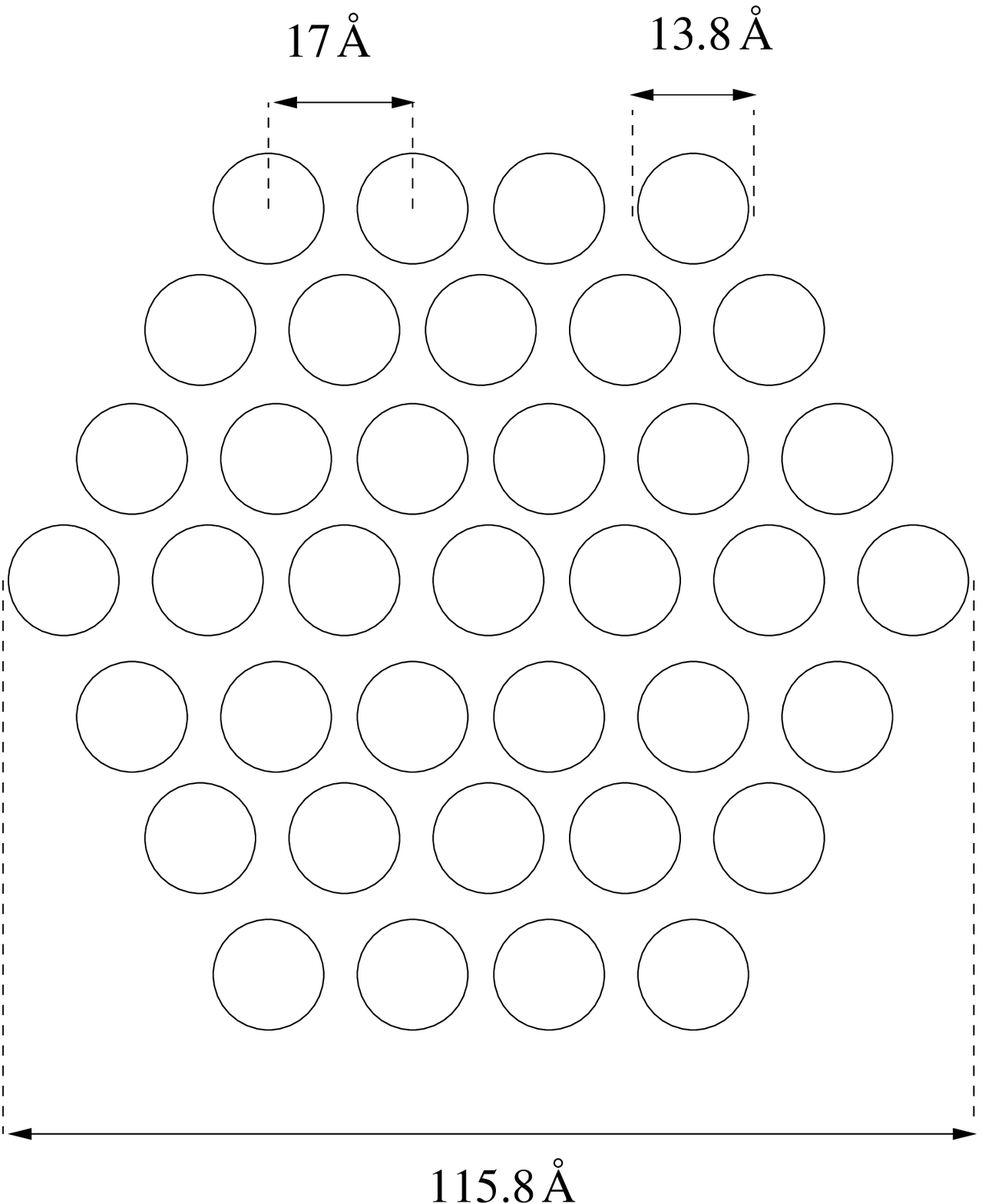}
\end{figure}
\begin{center}
 {\bf FIG. \protect\ref{fig:bundle}}
\end{center}

\newpage
\begin{figure}[ht]
\epsfysize=8.in \epsfbox{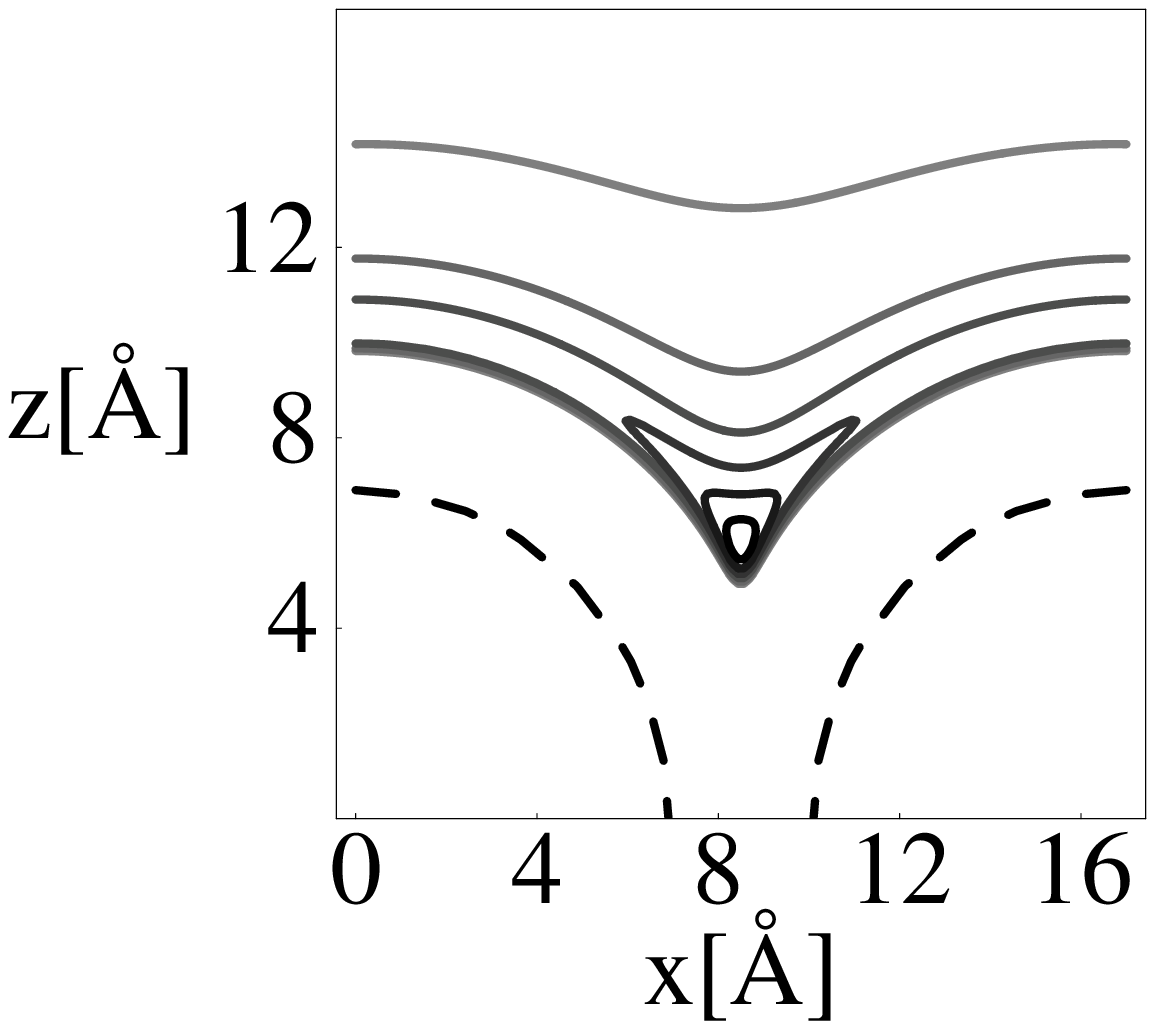}
\end{figure}
\begin{center}
 {\bf FIG. \protect\ref{fig:potential}}
\end{center}

\newpage
\begin{figure}[ht]
\epsfysize=5.in\epsfbox{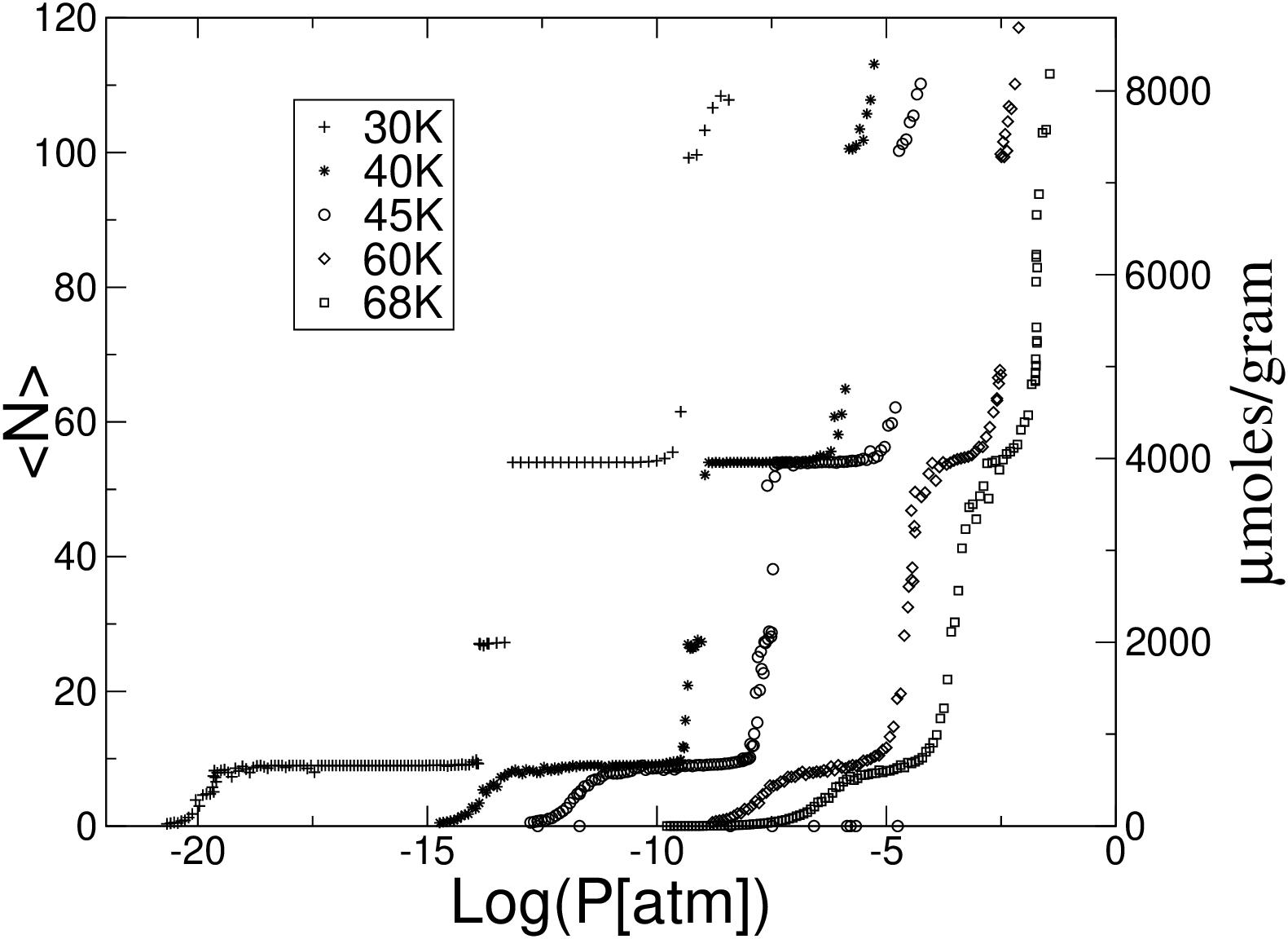}
\end{figure}
\vspace{5cm}
\begin{center}
 {\bf FIG. \protect\ref{fig:isoar}}
\end{center}

\newpage
\begin{figure}[ht]
\epsfysize=8.in \epsfbox{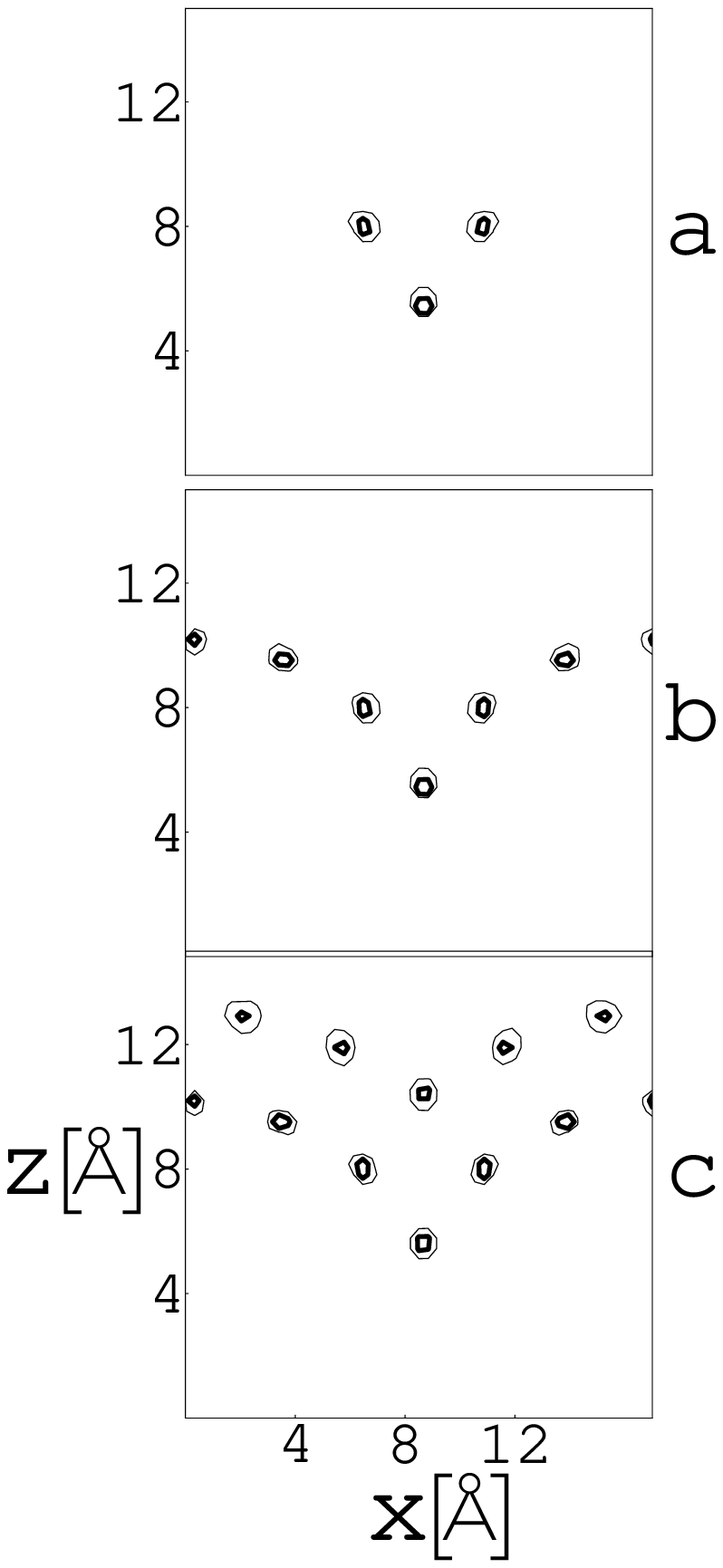}
\end{figure}
\begin{center}
 {\bf FIG. \protect\ref{fig:dens}}
\end{center}

\newpage
\begin{figure}[ht]
\epsfysize=5.in \epsfbox{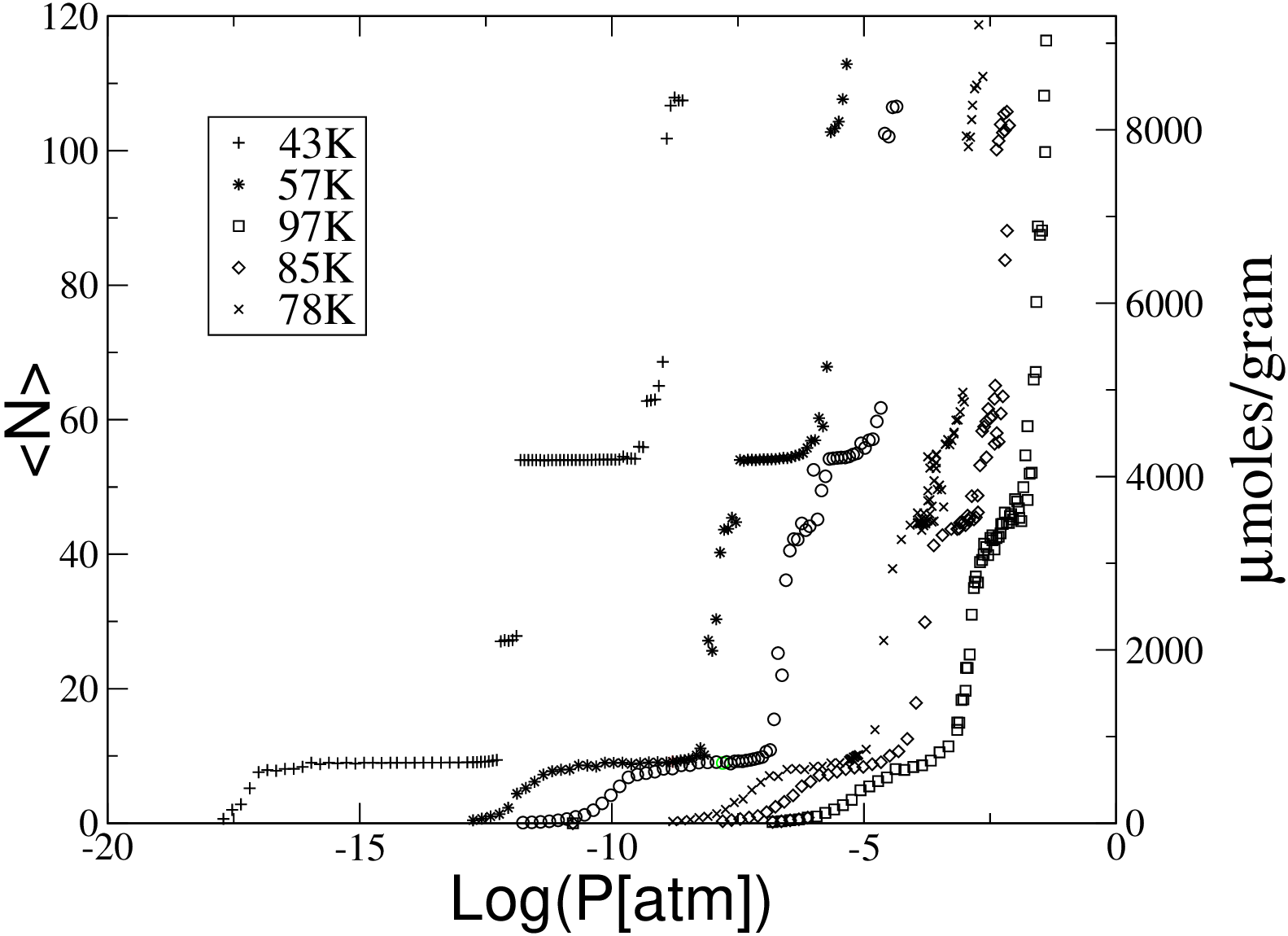}
\end{figure}
\vspace{5cm}
\begin{center}
 {\bf FIG. \protect\ref{fig:isokr}}
\end{center}

\newpage
\begin{figure}[ht]
\epsfysize=5.in \epsfbox{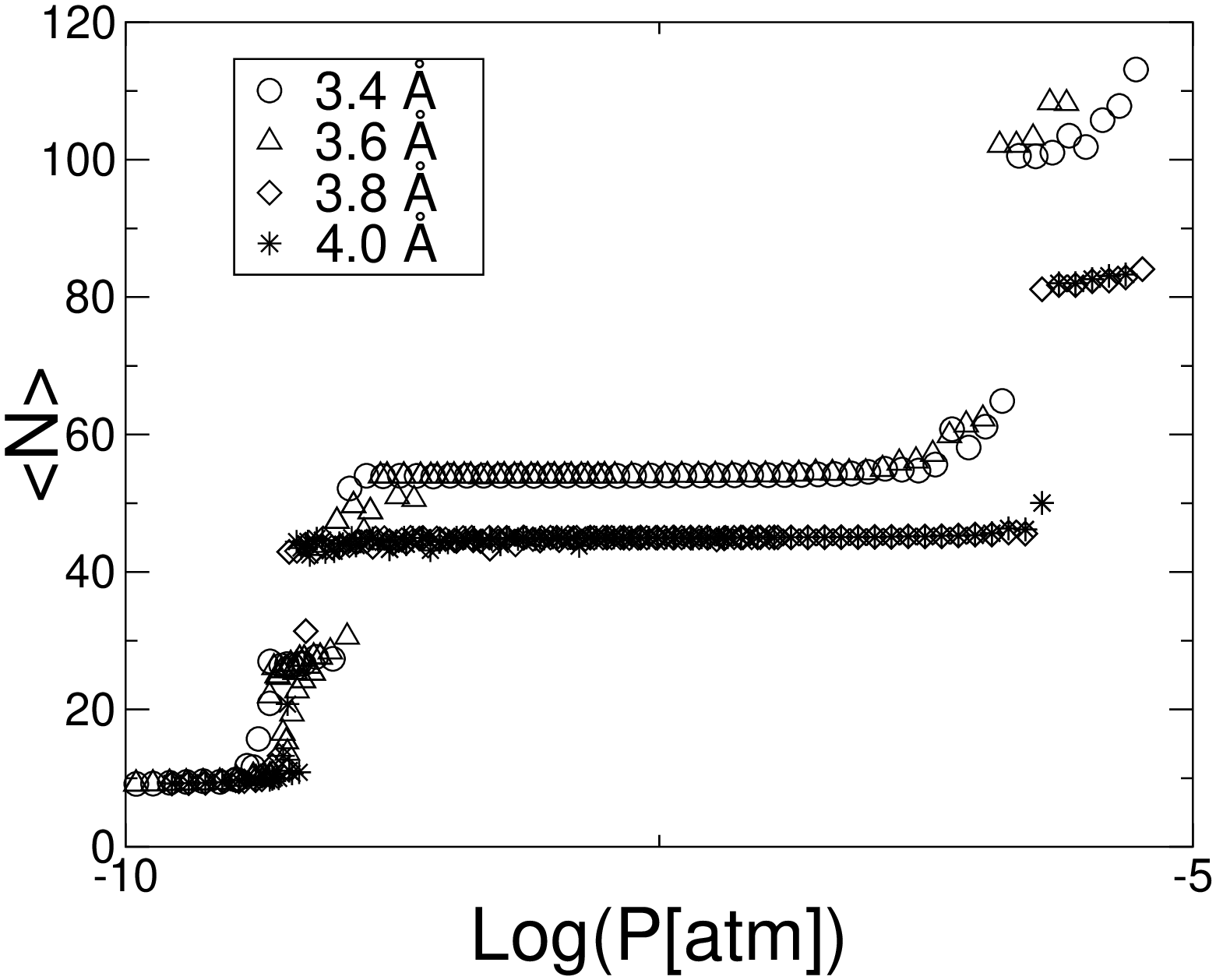}
\end{figure}
\vspace{5cm}
\begin{center}
 {\bf FIG. \protect\ref{fig:arti}}
\end{center}

\end{document}